\documentclass[prd, preprint, nofootinbib]{revtex4}

\usepackage{graphicx}
\usepackage{amsfonts}
\usepackage{latexsym}
\usepackage{amsmath}
\usepackage{amssymb}
\usepackage{epsfig}

\begin{document}

\vspace*{-2cm}

\title{Inelastic extra $U(1)$ charged scalar dark matter 
}

\author{Nobuchika Okada}
 \email{okadan@ua.edu}
 \affiliation{
Department of Physics and Astronomy, 
University of Alabama, Tuscaloosa, Alabama 35487, USA
}

\author{Osamu Seto}
 \email{seto@particle.sci.hokudai.ac.jp}
 \affiliation{Institute for the Advancement of Higher Education, Hokkaido University, Sapporo 060-0817, Japan}
 \affiliation{Department of Physics, Hokkaido University, Sapporo 060-0810, Japan}

%

\begin{abstract}
The null results in dark matter direct detection experiments imply the present scalar dark matter (DM)
 annihilation cross section to bottom quark pairs through the Higgs boson exchange is smaller than
 about $10^{-31}$ cm$^3/$s for a wide DM mass range, which is much smaller than the required annihilation
 cross section for thermal relic DM. 
We propose models of a thermal relic DM with the present annihilation cross section being very suppressed. 
This property can be realized in an extra $U(1)$ gauge interacting complex scalar DM,
 where the thermal DM abundance is determined by coannihilation through the gauge interaction
 while the present annihilation is governed by Higgs bosons exchange processes. 
An interaction between DM and the extra $U(1)$ breaking Higgs field generates
 a small mass splitting between DM and its coannihilating partner so that
 coannihilation becomes possible and also
 the $Z'$-mediated scattering off with a nucleon in direct DM search becomes inelastic.
We consider scalar dark matter in $U(1)_{B-L}, U(1)_{(B-L)_3}$ and $U(1)_{L_\mu-L_\tau}$ extended models
 and identify viable parameter regions. 
We also discuss various implications to future DM detection experiments,
 the DM interpretation of the gamma-ray excess in the globular cluster 47 Tucanae,
 the muon anomalous magnetic moment, the Hubble tension and others.
\end{abstract}

\preprint{EPHOU-19-011} 

\vspace*{3cm}

\maketitle

\section{Introduction}

Weakly interacting massive particle (WIMP) is a primary candidate for the dark matter (DM)
 in the Universe. 
An appealing property of WIMP is its complementarity: If WIMP is a thermal relic from
 the early Universe, the abundance of DM is determined by the annihilation cross section
 of the order of picobarn for WIMP annihilation into standard model (SM) particles
 at thermal freeze-out. 
This implies that WIMP annihilation occurs even today in a DM dense region with a similar
 magnitude of annihilation cross section, resulting in an excess
 of spectrum in various cosmic rays, such as gamma rays, neutrinos and charged particles. 
The Fermi-LAT has published limits on the DM annihilation cross section into final
 states generating gamma rays~\cite{Ackermann:2015zua,Fermi-LAT:2016uux}. 
Those limits have been expressed, in particular, annihilation modes into $b\bar{b}$ pair
 and $\tau^-\tau^+$ pair, because these could be dominant annihilation modes in a large
 class of WIMP models. 
In fact, for a WIMP mass smaller than about $100$ GeV, the obtained upper bound on the
 cross section is found to be smaller than that required for the thermal DM abundance of
 about $1$ pb~\cite{Ackermann:2015zua,Fermi-LAT:2016uux}.

Interactions of WIMP are constrained by the so-called direct DM detection experiments as well,
 through scattering processes between DM and a nucleon. 
Through those scatterings, WIMPs in our galactic halo are experimentally detectable. 
Various direct DM detection experiments such as the LUX~\cite{Akerib:2017kat},
 PandaX-II~\cite{Cui:2017nnn}, and XENON1T~\cite{Aprile:2018dbl} have not detected a
 significant signal, which sets the upper bound on the cross section to be smaller than
 $\mathcal{O}(10^{-9})$ pb for the wide range of DM mass. 
One may recognize a large hierarchical discrepancy of magnitude between the cross section
 for direct DM detection and that for annihilation.

As we will explicitly show in the next section, a theoretical interpretation of the small
 scattering cross section of WIMP with a nucleon implies that the present DM annihilation cross
 section for most of the DM mass range seems smaller than about $10^{-31}$ cm$^3/$s which is
 far below the sensitivity of the current and near-future observations. 
This would lead to incompatibility between the thermal DM abundance and such too small
 present DM annihilation cross sections. 
Thus, null detection of WIMP seems to confront with the desired annihilation cross section
 for thermal freeze-out.

Before we proceed discussion, we note several ways out of the above argument. 
The first is based on our assumption that the relevant scalar-type operator, $\chi\chi\bar{q}q$, between
 WIMP $\chi$ and quarks $q$ is present and unsuppressed. 
We usually expect that this operator is obtained after integrating out the Higgs boson in the SM. 
That argument is robust as long as the SM Higgs boson is a mediator. 
An exception is the case that WIMP and a mediator couple with not quarks but leptons only,
 and thus the operator $\chi\chi\bar{q}q$ does not exist. 
See, for example, Refs.~\cite{Krauss:2002px,Cheung:2004xm}.
Another case is that the operator is absent or very suppressed because WIMP-nucleon
 scattering processes occur in nonrelativistic regime. 
A pseudoscalar mediator dark matter~\cite{Ipek:2014gua} or pseudo-Nambu-Goldstone boson
 dark matter~\cite{Gross:2017dan} is such an example. 
The scattering cross section with a nucleon for these DMs has been studied in detail in
 Refs.~\cite{Arcadi:2017wqi,Pani:2017qyd,Sanderson:2018lmj,Li:2018qip,Abe:2018emu} and
 Refs.~\cite{Azevedo:2018exj,Ishiwata:2018sdi,Huitu:2018gbc,Alanne:2018zjm,Karamitros:2019ewv},
 respectively. 
Note that only WIMP-nucleon scattering cross section is suppressed,
 while WIMP annihilation cross sections both today and  in the early Universe
 can be about $1$ pb as usual.
Provided that the SM Higgs boson is a mediator, the second is based on the assumption
 that the annihilation process for freeze-out is same as that at present and kinematically
 $s$-wave. 
If an annihilation process for freeze-out is $p$-wave dominated, the present annihilation cross section
 is suppressed by the velocity squared $v^2 \sim 10^{-6}$ while that in the early Universe is not
 suppressed because the relative velocity is not so small as $v \sim 0.1$. 
A similar but moderated suppression can occur in the vicinity of the resonance pole due to a difference between
 relative velocities at present and in the early Universe. 
This was utilized in Refs.~\cite{Okada:2013bna,Okada:2014usa,Das:2016fwl} 
 to account for the Galactic Center gamma-ray excess~\cite{Goodenough:2009gk,Hooper:2010mq,Hooper:2011ti}. 
Thus, for example, if the freeze-out annihilation mode is $p$-wave and the present annihilation
 is dominated by another mode of $s$-wave (e.g., as in Ref.~\cite{Choi:2012ap}), the scenario is still consistent. 
There is yet another famous mechanism of different annihilation modes,
 that is coannihilation at freeze-out. 
The neutralino-stau coannihilation in supersymmetric models is a well-known example.

This paper is organized as follows.
In the next section, we show that the current constraints by the null result of DM direct detection experiments 
 imply that the present annihilation cross section into $b\bar{b}$ is typically smaller than
 $\mathcal{O}(10^{-31})$ cm$^3/$s. 
In Sec.~\ref{Sec:Models}, we introduce generic models for an extra $U(1)$ gauge interacting scalar DM,
 where the scalar DM with the extra $U(1)$ breaking Higgs field is also introduced,
 as a preparation for discussion based on specific extra $U(1)$ models in the following sections.
In the models, thermal DM abundance is determined by coannihilation through the gauge interaction
 while the present annihilation is governed by Higgs bosons exchange processes. 
We investigate three specific $U(1)$ models: $U(1)_{B-L}$ in Sec.~\ref{Sec:BL}, $U(1)_{(B-L)_3}$ in Sec.~\ref{Sec:BL3}
 and $U(1)_{L_\mu-L_\tau}$ in Sec.~\ref{Sec:MuTau}, respectively.
Section~\ref{Sec:Summary} is devoted to summary.

\section{Comparison of indirect and direct bounds}

\subsection{Dark matter elastic scattering with nuclei}

The spin-independent (SI) DM scattering cross section with nucleus ($N$)
 made of $Z$ protons ($p$) and $A-Z$ neutrons ($n$) 
 is given by~\cite{Jungman:1995df}
\begin{equation}
\sigma_{\rm SI}^N = \frac{1}{\pi}
 \left(\frac{m_N }{m_N + m_S}\right)^2 ( Z f_p + (A-Z) f_n )^2, 
\label{sigmaSI:scalar}
\end{equation}
 for a real scalar DM $S$ with the mass $m_S$.
The effective coupling with a proton $f_p$ and a neutron $f_n$ is expressed,
 by use of the hadronic matrix element, as
\begin{equation}
 \frac{f_i}{m_i} = \sum_{q=u,d,s}f_{Tq}^{(i)}\frac{\alpha_q}{m_q} 
  + \frac{2}{27}f_{TG}^{(i)}\sum_{c,b,t}\frac{\alpha_q}{m_q},
\end{equation}
 where $i = p, n$, and $\alpha_q$ is an effective coupling of the DM particle with a $q$-flavor quark 
 in the effective operator,
\begin{equation}
 \mathcal{L}\supset \alpha_q \bar{q}q S^2 ,
\label{coupling:Lqs}
\end{equation}
which is obtained by integrating out mediator particles from the original Lagrangian. 
For the origin of such an operator, we consider the scalar interaction terms and Yukawa interactions
 with quarks\footnote{
For illustrative purpose and simplicity, we ignore the possible contribution from other possible
 mediators such as scalar quarks in supersymmetric models. 
Note that a contribution from such particles is likely to be negligibly small compared with
 that by the Higgs boson exchange.}, 
\begin{equation}
\mathcal {L} \supset  - \lambda_4 v h \frac{S^2}{2} - \frac{m_q}{v} h \bar{q} q ,
\label{coupling:h}
\end{equation}
 where $v \simeq 246$ GeV is the vacuum expectation value (VEV) of the SM Higgs field
 and $h$ is the SM Higgs boson with the mass of $m_h \simeq 125$ GeV.

\subsection{Annihilation cross section for indirect signal}

The present annihilation cross section $(\sigma v)_0$ of the scalar DM particle ($S$)
 is given by its $s$-wave component
 of the annihilation cross section, e.g., by the limit of $v \rightarrow 0$.
From the interaction in Eq.~(\ref{coupling:h}) relevant to the direct detection,
 we also obtain the present day annihilation cross section to $b\bar{b}$ as
\begin{equation}
(\sigma v)_0
   = \frac{12 m_b^2}{m_S^2} \left|\frac{\lambda_4}{4 m_S^2-m_h^2+i m_h \Gamma_h}\right|^2 (m_S^2- m_b^2) ,
\label{annihilation:bbarb}
\end{equation}
 where $m_b$ and $\Gamma_h$ denote the mass of bottom quark and the total decay width of $h$, respectively.

\subsection{Comparison of the Fermi bound and direct search bounds}

Fermi-LAT Collaboration has shown that WIMP annihilation cross section to $b\bar{b}$ 
 should be less than $10^{-26} \mathrm{cm}{}^3/\mathrm{s}$
 for the WIMP mass $\lesssim 100$ GeV~\cite{Ackermann:2015zua,Fermi-LAT:2016uux}. 
Considering the fact that both the WIMP annihilation to $b\bar{b}$ (\ref{annihilation:bbarb})
 and the WIMP scattering with a nucleon (\ref{sigmaSI:scalar}) originate from the same interactions
 (\ref{coupling:h}), one may expect that the bound from direct DM search experiments
also sets a severe upper limit on the WIMP annihilation cross section.

\begin{figure}[htbp]
\centering
\includegraphics[clip,width=12.0cm]{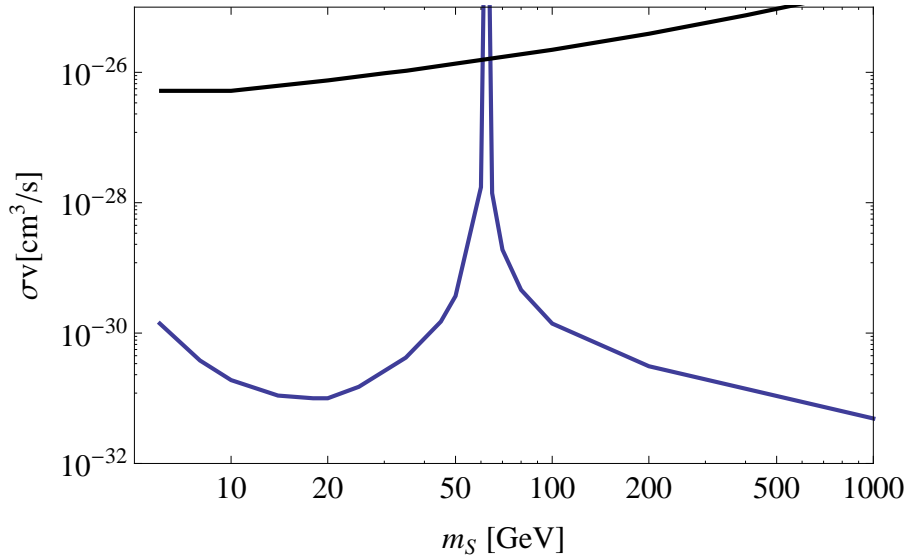}
\caption{
The upper bounds on WIMP annihilation cross section into $b\bar{b}$ are shown.
Fermi-LAT bound~\cite{Ackermann:2015zua,Fermi-LAT:2016uux}
 is shown as the black line. The blue curve represents the theoretical
interpretation of XENON1T bound into the upper bound for WIMP annihilation.
}
\label{Fig:sigmav:indirect}
\end{figure}

We show in Fig.~\ref{Fig:sigmav:indirect} the upper bound on the WIMP annihilation cross section
 into $b\bar{b}$ from the Fermi-LAT bound~\cite{Ackermann:2015zua,Fermi-LAT:2016uux}
(black solid line) and the theoretical interpretation of
XENON1T null result (blue curve). 
It is clear that the limit on the annihilation cross section
 derived from the direct DM detection bound is more stringent than the constraints on
 the annihilation cross section reported by the Fermi-LAT. We find the annihilation
 cross section is smaller than $\mathcal{O}(10^{-31} ) \mathrm{cm}{}^3/\mathrm{s}$
 in a wide range of the WIMP mass, except for the vicinity of $m_h/2$,
 where the DM pair-annihilation is enhanced by the Higgs boson resonance.\footnote{
As we have declared, we take account of only Higgs boson(s) exchange processes in this paper. 
However, we will not consider Majorana fermion DM because its annihilation processes
 through $s$-channel Higgs boson exchange is $p$-wave and suppressed by its velocity squared $v^2 \sim 10^{-6}$. 
For an explicit calculation, see Refs.~\cite{McDonald:2008up,Okada:2010wd,Djouadi:2011aa} for example.
}
%

Naive expectation based on the above consideration is that little cosmic gamma-ray excess
 can be generated by WIMP annihilation whose cross section should be very small due to
 the null results in direct DM detection experiments. 
One might expect another mediator Higgs boson could relax the bound
 as the SM Higgs boson does at the WIMP mass around $62$ GeV.
In Refs.~\cite{Okada:2013bna,Okada:2014usa,Das:2016fwl}, we have studied this possibility 
 in the context of Galactic Center excess~\cite{Goodenough:2009gk,Hooper:2010mq,Hooper:2011ti}
 and shown that not only can the predicted annihilation cross section not be suppressed
 by orders of magnitude but also the second mediator in type-II two Higgs doublet model
 confronts with the LHC results~\cite{Das:2016fwl}.

\section{Model}
\label{Sec:Models}

A consequence of the above discussion is how those constraints
 can be compatible with the annihilation cross section of the order of $10^{-26}$ cm$^3/$s
 for thermal freeze-out in the early Universe.
The present annihilation cross section $(\sigma v)_0$ of a WIMP,
 which is relevant for the indirect DM detection, is given by its $s$-wave component 
 of the self-pair annihilation cross section, i.e., by the limit of $v \rightarrow 0$. 
Note that in general the thermal averaged cross section $\langle\sigma v\rangle$ at the early Universe
 required to reproduce $\Omega h^2 \simeq 0.1$~\cite{Aghanim:2018eyx} is not necessarily the same as 
 the present annihilation cross section $(\sigma v)_0$. 

In the following, we will show that an extra $U(1)$ gauge interaction and its breaking
 provide large enough annihilation cross section for thermal freeze-out only and
 do not induce present annihilation processes much. 
Then, not only the observed $\Omega h^2$ can be reproduced but also 
 the expected WIMP-nucleon scattering cross section and
 the present WIMP pair annihilation cross section is small enough
 to be consistent with the null results in those DM search experiments.
In this section, we summarize the general aspect of scalar DM interacting through
 an extra $U(1)$ interaction before we discuss the details for specific $U(1)$ models
 in the following sections.

\subsection{Gauged $U(1)$ models}

\begin{table}[t]
\begin{center}
\begin{tabular}{|c|ccc|c|}
\hline
      &  SU(3)$_c$  & SU(2)$_L$ & U(1)$_Y$ & U(1)  \\ 
\hline
$Q^{i}$ & {\bf 3 }        &  {\bf 2}         & $ 1/6$     &  $q_{Q^i} $  \\
$u^{i}_{R}$ & {\bf 3 }    &  {\bf 1}         & $ 2/3$     &  $q_{u^i} $  \\
$d^{i}_{R}$ & {\bf 3 }    &  {\bf 1}         & $-1/3$     &  $q_{d^i} $  \\
\hline
$L^{i}$        & {\bf 1 }    &  {\bf 2}       & $-1/2$     & $q_{L^i} $  \\
$e^{i}_{R}$    & {\bf 1 }    &  {\bf 1}       & $-1$       & $q_{e^i} $  \\
\hline
$\Phi$            & {\bf 1 }    &  {\bf 2}       & $ 1/2$    & $0 $   \\  
\hline
$N^{i}_{R}$    & {\bf 1 }    &  {\bf 1}       &$0$         & $ q_{N^i} $   \\
$\phi_1$       & {\bf 1 }    &  {\bf 1}       &$ 0$        & $ + 1 $  \\
$\phi_2$       & {\bf 1 }    &  {\bf 1}       &$ 0$        & $ + 2 $  \\ 
\hline
\end{tabular}
\end{center}
\caption{
The particle content of our $U(1)$ model. 
In addition to the SM particle content of three generations ($i=1,2,3$),
 right-handed neutrinos ($N_R$) and two $U(1)$ Higgs fields ($\phi_1$ and $\phi_2$) are introduced.
Charges of fermions depend on the specific gauge group and are assigned to make the model free from anomalies.
}
\label{table1}
\end{table}

We consider an extend SM based on the gauge group $SU(3)_C \times SU(2)_L \times U(1)_Y \times U(1)$. 
In addition to the SM model particles, we introduce right-handed neutrinos ($N_R$)
 and two SM singlet Higgs fields ($\phi_1$ and $\phi_2$) charged under the new extra $U(1)$ gauge symmetry.
We identify $\phi_1$ with a DM candidate while $\phi_2$ is a Higgs field responsible to break
 the extra $U(1)$ gauge symmetry. 
Here, the subscripts 1 and 2 stand for those gauge charges.

The particle content is listed on Table~\ref{table1}.
Charges of fermions under the specific gauge group must be assigned to make the model free from anomalies.
The scalar potential is expressed as~\cite{Chao:2017ilw,Okada:2018xdh}
\begin{align}
V(\Phi, \phi_1, \phi_2 )
 =& - M^2_{\Phi} |\Phi|^2 + \frac{\lambda}{2} |\Phi|^4 + M^2_{\phi_1} \phi_1\phi_1^{\dagger} - M^2_{\phi_2} \phi_2\phi_2^{\dagger}   \nonumber \\
 & + \frac{1}{2} \lambda_1 (\phi_1 \phi_1^{\dagger})^2+\frac{1}{2}\lambda_2 (\phi_2\phi_2^{\dagger} )^2
 +\lambda_3 \phi_1\phi_1^{\dagger} (\phi_2 \phi_2^{\dagger}) \nonumber \\
 & +  (\lambda_4 \phi_1\phi_1^{\dagger} + \lambda_5 \phi_2\phi_2^{\dagger})|\Phi|^2
 - A (\phi_1 \phi_1 \phi_2^{\dagger} + \phi_1^{\dagger} \phi_1^{\dagger} \phi_2 ) ,
\label{eq:totalpotential}
\end{align}
with $\Phi$ being the SM Higgs field.
All parameters in the potential~(\ref{eq:totalpotential}) are taken to be real and positive.

Here, we emphasize that, in previous works on an extra gauged $U(1)$ charged
 scalar DM~\cite{Rodejohann:2015lca,Biswas:2016ewm,Singirala:2017see,Bandyopadhyay:2017bgh,Biswas:2016yan,Garani:2019fpa},
 the scalar DM does not interact with the $U(1)$ breaking Higgs field
 through a trilinear coupling due to its gauge charge,
 and the $\phi_1 \phi_1 \phi_2^{\dagger}$ corresponding term
 in our scalar potential (\ref{eq:totalpotential}) is absent.
On the other hand, the presence and the effect of this term is essential in this paper,
 as we will show below.

\subsection{Dark matter mass and interactions}

At the $U(1)$ and the electroweak (EW) symmetry breaking vacuum, the SM Higgs field and
 the $U(1)$ Higgs field are expanded around those VEVs, $v$ and $v_2$, as
\begin{align}
\Phi =& \left( \begin{array}{c}
          0 \\
          \frac{v + \varphi}{\sqrt{2}} \\
         \end{array}
  \right), \\
\phi_1 =& \frac{ S + i P }{\sqrt{2}} ,\\
\phi_2 =& \frac{ v_2 + \varphi_2 }{\sqrt{2}} .
\end{align}
 where those VEVs are determined by the stationary conditions:
\begin{align}
 -M^2_{\Phi} + \frac{1}{2}\lambda v^2+\frac{1}{2}\lambda_5 v_2^2 = 0, 
\label{eq:tadpole1} \\
 -M^2_{\phi_2} + \frac{1}{2}\lambda_2 v_2^2+\frac{1}{2}\lambda_5 v^2 = 0 .
\label{eq:tadpole2}
\end{align}
Mass terms of particles are expressed as
\begin{align}
\mathcal{L}_\mathrm{mass} = & -\frac{1}{2}(\varphi \,\, \varphi_2  )
\left(
\begin{array}{cc}
 -M^2_{\Phi} + \frac{3}{2} \lambda v^2 + \frac{1}{2}\lambda_5 v_2^2 & \lambda_5 v v_2 \\
 \lambda_5 v v_2 & -M^2_{\phi_2} + \frac{3}{2} \lambda_2 v_2^2 + \frac{1}{2}\lambda_5 v^2  \\
\end{array}
\right)
\left(
\begin{array}{c}
 \varphi \\
 \varphi_2 \\
\end{array}
\right)   \nonumber \\ 
& - \frac{1}{2}
\left(
 M_{\phi_1}^2+\frac{1}{2}\lambda_3v_2^2+\frac{1}{2}\lambda_4 v^2-\sqrt{2}A v_2
\right)S^2   \nonumber \\
& - \frac{1}{2}
\left(
 M_{\phi_1}^2+\frac{1}{2}\lambda_3v_2^2+\frac{1}{2}\lambda_4 v^2+\sqrt{2}A v_2
\right)P^2   \nonumber \\
 & - \frac{1}{2} g^\prime{}^2 4 v_2^2 Z^{\prime \mu} Z^\prime_\mu ,
\end{align}
 where $g^\prime$ is the $U(1)$ gauge coupling.
The physical states ($\varphi$ and $\varphi_2$) are diagonalized 
 to the mass eigenstates ($h$ and $H$) with masses $m_h$ and $m_H$ as 
 \begin{equation}
  \left( \begin{array}{c}
          \varphi  \\
          \varphi_2  \\
         \end{array}\right)
   = 
  \left( \begin{array}{cc}
          \cos\alpha & \sin\alpha\\
          -\sin\alpha & \cos\alpha\\
         \end{array}\right)
  \left( \begin{array}{c}
          h \\
          H \\
         \end{array}\right) .
 \end{equation}
For a small mixing angle $\alpha$, $h$ is identified with the SM-like Higgs boson.
In fact, we will take $\alpha \simeq 0.001$ in the following analysis.
With the $U(1)$ and the EW symmetry breaking, the $Z^\prime$ boson,
 $S$ and $P$ acquire their masses, respectively, as 
\begin{align}
  m_{Z^\prime}^2 =& g^\prime{}^2 4v_2^2, \\
  m_S^2 =& M_{\phi_1}^2+\frac{1}{2}\lambda_3v_2^2+\frac{1}{2}\lambda_4 v^2-\sqrt{2}A v_2, \\
  m_P^2 =& M_{\phi_1}^2+\frac{1}{2}\lambda_3v_2^2+\frac{1}{2}\lambda_4 v^2+\sqrt{2}A v_2 .
\end{align}
Note that the parameter $A$ controls the mass splitting between $S$ and $P$. 
Since we take $A$ positive, $S$ is lighter than $P$ and becomes the DM candidate.
Here, we note that
 the mass degeneracy of coannihilating particles are accidental
 in many cases of coannihilation including that in supersymmetric models.
On the other hand, the mass degeneracy between $S$ and $P$ in our model
 would be reasonable because both are originally in the same multiplet with the common mass. 

Three point interaction terms among $S (P)$ and $h (H)$ are expressed as
\begin{align}
\mathcal{L}_\mathrm{int} \supset & \frac{1}{2}\left( \left(\lambda_4 v \cos\alpha  -(\lambda_3 v_2-\sqrt{2}A)\sin\alpha \right) h + \left( \lambda_4 v \sin\alpha +(\lambda_3 v_2-\sqrt{2}A) \cos\alpha \right) H  \right) S^2 \nonumber\\
& + \frac{1}{2}\left( \left( \lambda_4 v \cos\alpha -(\lambda_3 v_2+\sqrt{2}A)\sin\alpha \right) h +  \left(\lambda_4 v \sin\alpha +(\lambda_3 v_2+\sqrt{2}A) \cos\alpha \right) H \right) P^2 .
\label{eq:Lag:Higgs-DM-DM}
\end{align}
The Yukawa interactions can then be written as
\begin{align}
 \mathcal{L}_\mathrm{Yukawa} \supset & 
  - \frac{m_{u^i} \cos\alpha}{v} h \bar u^i u^i - \frac{m_{u^i} \sin\alpha}{v} H \bar u^i u^i
 - \frac{m_{d^i} \cos\alpha}{v} h \bar d^i d^i  - \frac{m_{d^i}\sin\alpha}{v} H \bar d^i d^i   \nonumber \\
 & - \frac{m_{d^i} \cos\alpha}{v} h \bar \ell^i \ell^i - \frac{m_{d^i}\sin\alpha}{v} H \bar \ell^i \ell^i . 
 \end{align}
Gauge interaction of the DM particle is expressed as
\begin{equation}
\mathcal{L}_\mathrm{int} =  g'_{} Z'{}^{\mu}\left( (\partial_{\mu}S) P- S \partial_{\mu}P \right) ,
\label{eq:Lag:gauge-DM-DM}
\end{equation}
 and similarly all generation quarks and leptons also interact to $Z'$ with corresponding charges.
The absence of $Z'$-DM-DM coupling means that the $Z'$-mediating DM scattering off
 with a nucleon is inelastic and ineffective for the mass splitting larger than
 the energy transfer in the scatterings~\cite{Hall:1997ah,TuckerSmith:2001hy}.
Hence, the DM-nucleon scattering in our model is Higgs exchange dominated. 

We here count the number of additional free parameters in our model
 except for the SM gauge and Yukawa couplings.
One is the extra gauge coupling $g'$ and the scalar potential (\ref{eq:totalpotential})
 contains ten parameters. 
$\lambda_1$ in the potential~(\ref{eq:totalpotential}) controls only DM self-interaction
 and hence is irrelevant in our discussion.
The other two are fixed by the SM Higgs VEV $v=246$ GeV and its mass $m_h = 125$ GeV.
Practically, we have in total eight parameters: $g', m_{Z'}, m_S, m_P, \alpha, m_H,
 \lambda_3$ and $\lambda_4$.
In the following analysis, we fix
 $(m_P-m_S, \sin\alpha, \lambda_3, \lambda_4) = (0.01 m_S, 1\times 10^{-3}, 1\times 10^{-3}, 0)$
\footnote{We take as this in order to have an efficient coannihilation
 and to have a tiny effective coupling between $h$ and $S$,
 which can be seen from Eqs.~(\ref{coupling:Lqs}) and (\ref{annihilation:bbarb}).}
 and vary the following four parameters: $g', m_{Z'}, m_S$, and $m_H$.

\subsection{Thermal relic abundance}

We estimate the thermal relic abundance of the real scalar DM, $S$, by solving the Boltzmann equation,
\begin{equation}
 \frac{d n }{dt}+3H n =-\langle\sigma_\mathrm{eff} v\rangle ( n^2 - n_{\rm EQ}^2),
\label{eq:boltzman}
\end{equation}
 where $H$ and $n_{\rm EQ}$ are the Hubble parameter and the DM number density at thermal equilibrium,
 respectively~\cite{KolbTurner}.  
In our model, the main annihilation mode is coannihilation $S P \rightarrow f\bar{f}$ through $s$-channel $Z'$ exchange for $m_{Z'} > m_S$ and the annihilation mode $S S \rightarrow Z^\prime Z'$ by $u(t)$-channel $P$ exchange for $m_{Z'} < m_S$ (see Appendixes for the formula we employ in our analysis).
We use the effective thermal averaged annihilation cross section
\begin{equation}
 \langle\sigma_\mathrm{eff}v\rangle  = \sum_{i,j= S,P} \langle\sigma_{ij} v_{ij}\rangle\frac{n_i}{n_{\rm EQ}}\frac{n_j}{n_{\rm EQ}},
\end{equation}
 to include the coannihilation effects properly and $n$ in Eq.~(\ref{eq:boltzman}) should be understood as $n = \sum_i n_i$ for $i = S, P$~\cite{Griest:1990kh,Edsjo:1997bg}.

\subsection{Signal prospect for direct and indirect dark matter search experiments}

We use the formula (\ref{sigmaSI:scalar}) for the evaluation of the DM-nucleon scattering cross section
 in an extra $U(1)$ extended model,
 because the $Z^\prime$ boson does not contribute to the scattering process.

Annihilation cross section for indirect signal is given by its $s$-wave component 
 of the self-pair annihilation cross section. 
A pair of scalar DM particles dominantly annihilates into $b \bar{b}$ and the weak gauge bosons $W$ and $Z$
 through the $s$-channel exchange of the Higgs bosons ($h$ and $H$). 
The cross section is enhanced by $h(H)$-boson exchange for $m_{h(H)} \sim 2m_S$ and suppressed by
 the destructive interference between two.
In addition, for $m_{Z'} < m_S$, the annihilation mode $S S \rightarrow Z^\prime Z'$
 by $u(t)$-channel $P$ exchange opens and has significant $s$-wave component, which is expressed as
\begin{equation}
\sigma v (SS\rightarrow Z'Z')|_{s-\mathrm{wave}}
   = \frac{g^\prime{}^4}{\pi}\frac{m_S^2 \left(m_S^2-m_{Z'}^2\right)^2}{m_{Z'}^4 \left(m_{Z'}^2-2 m_S^2 \right)^2}
    \sqrt{1-\frac{m_{Z'}^2}{m_S^2}} ,
\label{annihilation0:ZpZp}
\end{equation}
 in the $m_P \rightarrow m_S$ limit.
Hence, the $m_{Z'} < m_S$ case is well constrained by Fermi-LAT or other DM indirect detection experiments.

\section{$B-L$ Model}
\label{Sec:BL}

\subsection{$B-L$ seesaw model}

\begin{table}[htbp]
\begin{center}
\begin{tabular}{|c|ccc|c|}
\hline
      &  SU(3)$_c$  & SU(2)$_L$ & U(1)$_Y$ & U(1)$_{B-L}$  \\ 
\hline
$Q^{i}$ & {\bf 3 }        &  {\bf 2}         & $ 1/6$     & $1/3 $   \\
$u^{i}_{R}$ & {\bf 3 }    &  {\bf 1}         & $ 2/3$     & $1/3 $   \\
$d^{i}_{R}$ & {\bf 3 }    &  {\bf 1}         & $-1/3$     & $1/3 $   \\
\hline
$L^{i}$        & {\bf 1 }    &  {\bf 2}       & $-1/2$     & $-1 $    \\
$e^{i}_{R}$    & {\bf 1 }    &  {\bf 1}       & $-1$       & $-1 $    \\
\hline
$\Phi$            & {\bf 1 }    &  {\bf 2}       & $ 1/2$    & $0 $   \\  
\hline
$N^{i}_{R}$    & {\bf 1 }    &  {\bf 1}       &$0$         & $-1 $    \\
$\phi_1$       & {\bf 1 }    &  {\bf 1}       &$ 0$        & $ + 1 $  \\
$\phi_2$       & {\bf 1 }    &  {\bf 1}       &$ 0$        & $ + 2 $  \\ 
\hline
\end{tabular}
\end{center}
\caption{
The particle content of our $U(1)_{B-L}$ model. 
In addition to the SM particle content ($i=1,2,3$), three RH neutrinos  
  ($N_R^i$ ($i=1, 2, 3$)) and two $U(1)_{B-L}$ Higgs fields ($\phi_1$ and $\phi_2$) are introduced.   
}
\label{tableBL}
\end{table}

First, we consider a model based on the gauge group $SU(3)_C \times SU(2)_L \times U(1)_Y \times U(1)_{B-L}$~\cite{Mohapatra:1980qe,Marshak:1979fm}.
In addition to the SM particle content with three generations ($i=1,2,3$), three RH neutrinos  
  ($N_R^i$ ($i=1, 2, 3$)) and two $U(1)_{B-L}$ Higgs fields ($\phi_1$ and $\phi_2$) are introduced. 
In the presence of the three RH neutrinos, the model is free from all the gauge and
 mixed gauge-gravitational anomalies.
The particles and those charges are listed on Table~\ref{tableBL}.

\subsection{Thermal relic abundance}

\begin{figure}[htbp]
\centering
\includegraphics[clip,width=11.0cm]{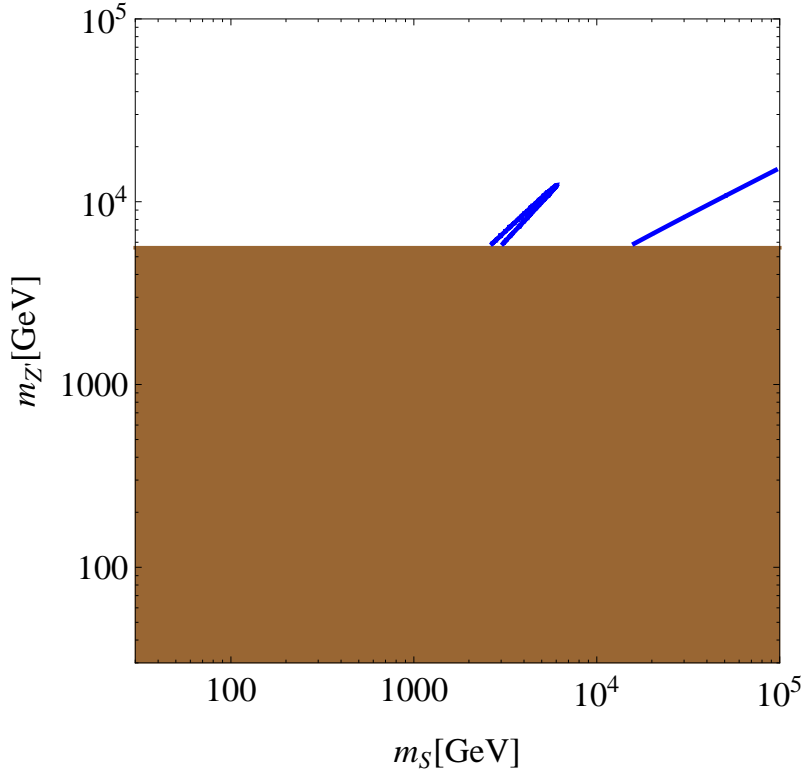}
\caption{
The contour (in blue) along which the observed DM relic abundance $\Omega h^2 \simeq 0.1$ is reproduced for $g_{B-L}=0.5$.
The excluded region for the $Z'$ boson mass by the LHC is brown shaded.
}
\label{Fig:blthermal}
\end{figure}

We show in Fig.~\ref{Fig:blthermal} the relation between $m_S$ and $m_{Z^\prime}$ drawn with blue lines
 in order to reproduce the observed DM relic abundance $\Omega h^2\simeq 0.1$~\cite{Aghanim:2018eyx}.
Since the so-called unitarity bound sets an upper limit on a WIMP mass to be less than a hundred TeV~\cite{Griest:1989wd},
 the curve is grown up to the WIMP mass of one hundred TeV.
The experimental bound on the mass of $Z^\prime$ in the $U(1)_{B-L}$ model has been derived based on
 the LEP and Tevarton data~\cite{Carena:2004xs} as well as
 the latest LHC Run 2 results~\cite{Amrith:2018yfb,Das:2019fee} that are expressed as a brown shaded region.
There are two cases reproducing the observed DM abundance:
 one is $m_S \sim$ several TeV by the resonant annihilation through $s$-channel $Z^\prime$ exchange and 
 the other is due to the $S S \rightarrow Z^\prime Z^\prime$ annihilation for $m_S > 10$ TeV.
For a smaller $g_{B-L}$, DM is overabundant and no solution to $\Omega h^2\simeq 0.1$ is found.
%

\subsection{Prospect for direct and indirect dark matter detection experiments}

\begin{figure}[htbp] 
 \centering
    \begin{tabular}{c}
 \begin{minipage}{0.48\hsize}
\centering
\includegraphics[width=7.5cm]{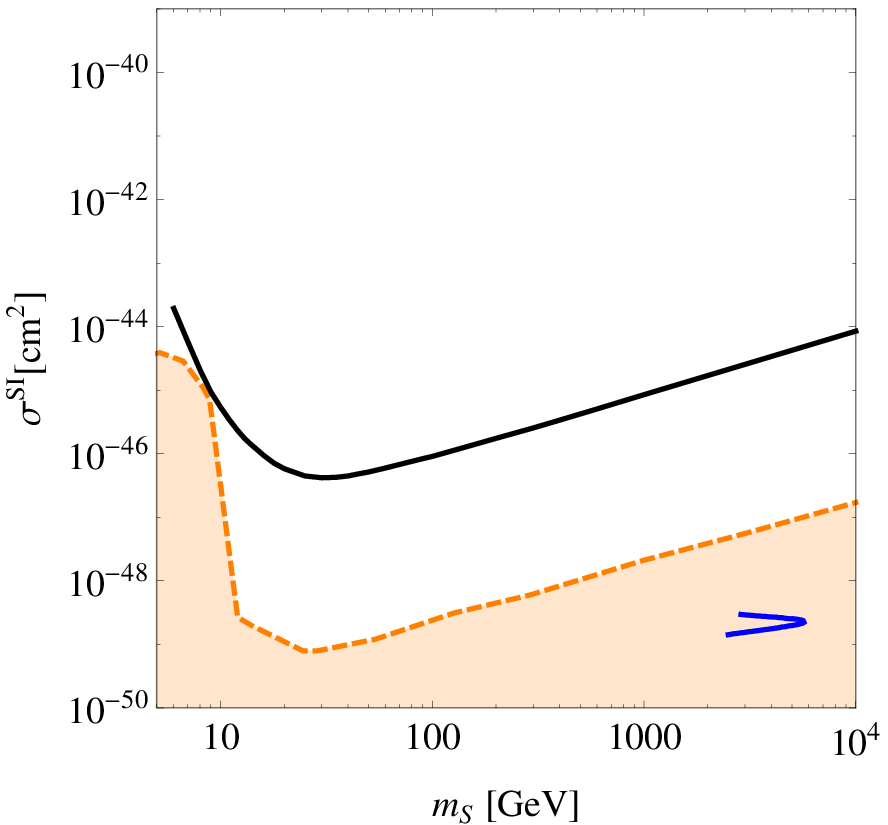}
 \end{minipage}
      \begin{minipage}{0.04\hsize}
        \hspace{2mm}
      \end{minipage}
 \begin{minipage}{0.48\hsize}
\centering
\includegraphics[width=7.5cm]{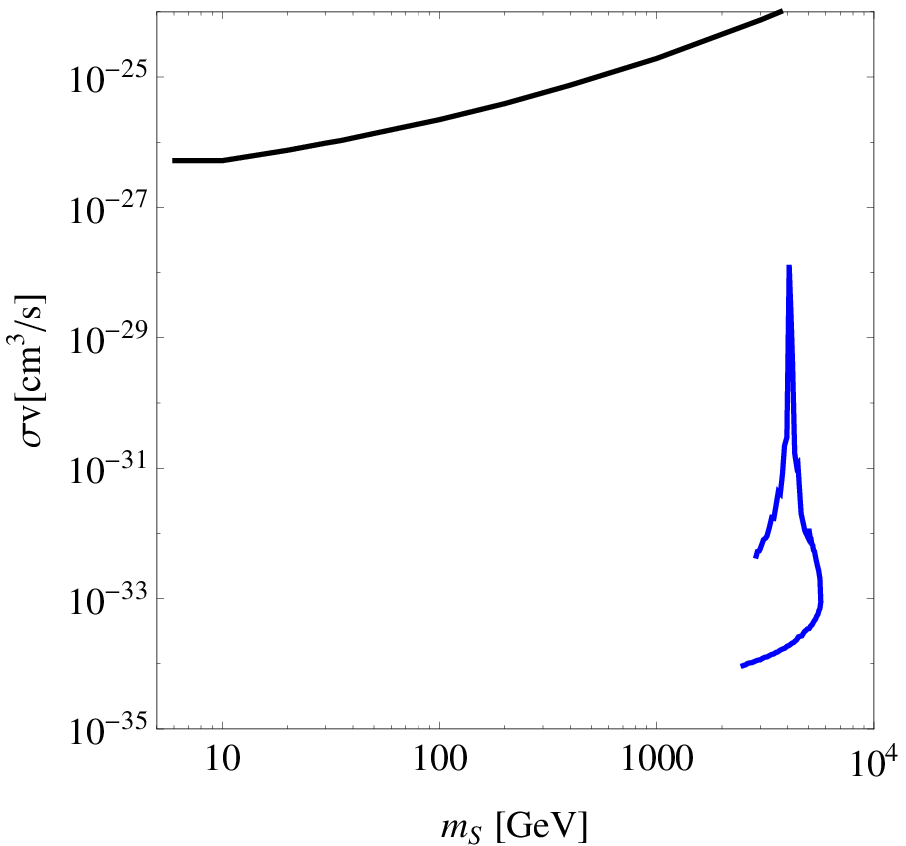}
 \end{minipage}   
 \end{tabular}
\caption{
Left panel: The prediction of the WIMP-nucleon scattering cross section (blue line) corresponding to Fig.~\ref{Fig:blthermal},
 along with the black solid curve being the upper bound obtained by the XENON1T(2018) and
 the orange shading with the dashed boundary curve for the neutrino background level. 
Right panel: The prediction of the present DM pair annihilation cross section (blue line) corresponding to Fig.~\ref{Fig:blthermal}
 along with the black solid curve being the upper bound obtained by Fermi-LAT Pass 8 data.
}
\label{Fig:bldetection}
\end{figure}

We show, in the left panel of Fig.~\ref{Fig:bldetection}, the spin-independent cross section,
 $\sigma^{\mathrm{SI}}$, as a function of the DM mass for the $Z^\prime$ funnel region in Fig.~\ref{Fig:blthermal},
 along with the black solid curve and the dashed orange curve being the XENON1T(2018) limit
 and the neutrino background level~\cite{Cushman:2013zza}, respectively.
It is clear that the predicted cross section lies below the neutrino background.
We show in the right panel of Fig.~\ref{Fig:bldetection} the prediction of the present DM pair annihilation cross section for the $Z^\prime$ funnel region in Fig.~\ref{Fig:blthermal}, along with the solid curve being the Fermi-LAT limit.
The sharp enhancement in the right panel is due to accidental resonance of the $H$ boson with the mass $m_H = v_2$.
%


\section{$(B-L)_3$ Model}
\label{Sec:BL3}

\subsection{$(B-L)_3$ seesaw model}

\begin{table}[t]
\begin{center}
\begin{tabular}{|c|ccc|c|}
\hline
      &  SU(3)$_c$  & SU(2)$_L$ & U(1)$_Y$ & U(1)$_{(B-L)_3}$  \\ 
\hline
$Q^{i}$ & {\bf 3 }        &  {\bf 2}         & $ 1/6$     & \, $0 $ for $i=1,2$  \quad $1/3 $ for $i=3$ \,  \\
$u^{i}_{R}$ & {\bf 3 }    &  {\bf 1}         & $ 2/3$     & \, $0 $ for $i=1,2$  \quad $1/3 $ for $i=3$ \,  \\
$d^{i}_{R}$ & {\bf 3 }    &  {\bf 1}         & $-1/3$     & \, $0 $ for $i=1,2$  \quad $1/3 $ for $i=3$ \,  \\
\hline
$L^{i}$        & {\bf 1 }    &  {\bf 2}       & $-1/2$     & \, $0 $ for $i=1,2$  \quad $-1 $ for $i=3$ \,   \\
$e^{i}_{R}$    & {\bf 1 }    &  {\bf 1}       & $-1$       & \, $0 $ for $i=1,2$  \quad $-1 $ for $i=3$ \,   \\
\hline
$\Phi$            & {\bf 1 }    &  {\bf 2}       & $ 1/2$    & $0 $   \\  
\hline
$N^{i}_{R}$    & {\bf 1 }    &  {\bf 1}       &$0$         & \, $0 $ for $i=1,2$  \quad $-1 $ for $i=3$ \,     \\
$\phi_1$       & {\bf 1 }    &  {\bf 1}       &$ 0$        & $ + 1 $  \\
$\phi_2$       & {\bf 1 }    &  {\bf 1}       &$ 0$        & $ + 2 $  \\ 
\hline
\end{tabular}
\end{center}
\caption{
The particle content of our $U(1)_{(B-L)_3}$ model. 
In addition to the SM particle content ($i=1,2,3$), three RH neutrinos  
  ($N_R^i$ ($i=1, 2, 3$)) and two $U(1)_{(B-L)_3}$ Higgs fields ($\phi_1$ and $\phi_2$) are introduced.   
}
\label{tableBL3}
\end{table}

Next, we consider a model based on the gauge group $SU(3)_C \times SU(2)_L \times U(1)_Y \times U(1)_{(B-L)_3}$. 
This ``flavored'' $B-L$ symmetry on the third generation is motivated by the fact that
 the anomaly cancellation of $U(1)_{B-L}$ gauge symmetry
 can be realized for each generation of fermions~\cite{Babu:2017olk,Alonso:2017uky,Bian:2017rpg,Cox:2017rgn}.
The particles and those charges are listed on Table~\ref{tableBL3}. Yukawa couplings are given by 
\begin{align}
\mathcal{L}_\mathrm{Yukawa} 
   =& \sum_{i,j = 1-3}\left( - y^{\ell}_{ij}  \overline{L^i} \Phi \ell_R^j
     - y^{u}_{ij}  \overline{Q^i} \widetilde{\Phi} u_R^j
     - y^{d}_{ij}  \overline{Q^i} \Phi d_R^j 
     - y^{D}_{ij} \overline{L^i} \widetilde{\Phi} N_R^j \right) \nonumber \\
 &    - \sum_{i = 1-2}\frac{1}{2} \overline{N_R^{i~C}} M_i N_R^i - \frac{1}{2} \overline{N_R^{3~C}} y_{N} \phi_2 N_R^3
           + \mathrm{ H.c.} ,
\label{Yukawa} 
\end{align}
 where $Q^i$ ($L^i$) is the ordinary left-handed quark (lepton) 
 in the $i$ th generation, $u_R^i$ and $d_R^i$ ($e_R^i$) are 
 the right-handed $SU(2)$ singlet up- and down-type quarks  (charged leptons).
However, due to  $U(1)_{(B-L)_3}$ gauge symmetry, $y_{i3}$ and $y_{3i}$ elements for $i=1, 2$ vanish, 
and thus realistic fermion flavor mixings cannot be reproduced.
To overcome this problem, a few successful UV completions have been proposed:
One is an extension of Higgs sector by Babu \textit{et al}. in Ref.~\cite{Babu:2017olk} 
 and another is introduction of heavy vectorlike fermions with additional scalars by Alonso \textit{et al}. in Ref.~\cite{Alonso:2017uky}. 
Since the details of those UV completions are irrelevant for our following discussion on DM phenomenology,
 we adopt, for simplicity, Lagrangian (\ref{Yukawa}) as an effective and relevant part of the full model.\footnote{
Successful analysis in Ref.~\cite{Cox:2017rgn} also shows that analysis based on a simplified Lagrangian works sufficiently well.}
In the model, the $U(1)_{(B-L)_3}$ gauge symmetry is broken by the VEV of $\phi_2$,
 the Majorana mass term of the third RH neutrino and the extra neutral gauge boson $Z'$ are generated. 

The experimental constraints on this model comes from lepton universality derived by
 the LEP~\cite{Chun:2018ibr} and the \textit{BABAR}~\cite{delAmoSanchez:2010bt} experiments.\footnote{
For a $Z^\prime$ boson search at the LHC, we refer the results with ditau final state~\cite{Aaboud:2017sjh}. 
Using a $Z^\prime$ boson production cross section from $b$ $\bar{b}$ annihilation presented 
 in, for example, Ref.~\cite{Faroughy:2016osc} and $BR( Z^\prime \to \tau^+ \tau^-) \sim 0.5$, 
 we conclude that the parameter set used in our analysis is consistent with the current LHC results. 
For future search reach with a large luminosity, see Ref.~\cite{Elahi:2019drj} and a part of parameter space of the $(B-L)_3$ model will be explored. 
}
In fact, the \textit{BABAR} experiment for testing a lepton universality in $\Upsilon(1S)$ decays resulted in the most stringent bound as
 $R_{\mu\tau}(\Upsilon)=1.005 \pm 0.013(\mathrm{stat.})\pm 0.022(\mathrm{syst.})$~\cite{delAmoSanchez:2010bt}.
Since the decay width ratio can be expressed as
\begin{equation}
R_{\mu\tau}(\Upsilon) \simeq \left( 1 + \frac{g_{(B-L)_3}^2}{e^2}\frac{m_{\Upsilon}^2}{m_{Z'}^2-m_{\Upsilon}^2}\right)^2 ,
\end{equation}
 where $m_{\Upsilon}$ is the mass of $\Upsilon$, we have constraints on $g_{(B-L)_3}$ and $m_{Z'}$.

\subsection{Thermal relic abundance}

\begin{figure}[htbp]
\centering
\includegraphics[clip,width=12.0cm]{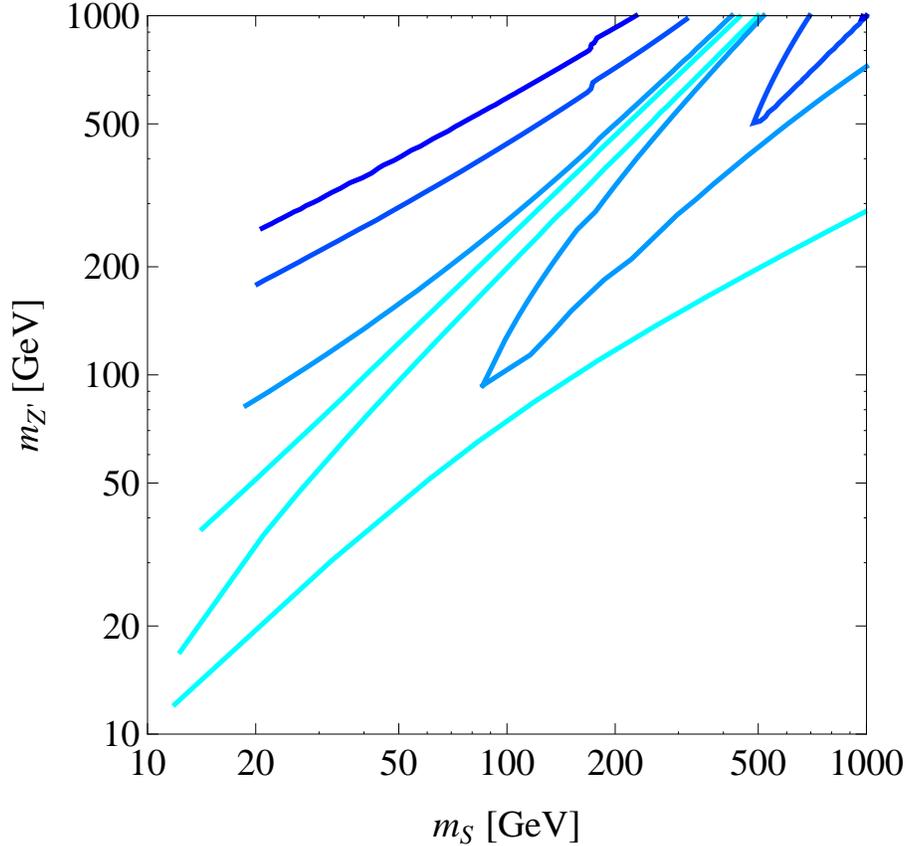}
\caption{
The relation between $m_S$ and $m_{Z'}$ to reproduce $\Omega h^2 \simeq 0.1$
 for various values of $g_{(B-L)_3} = 1$ (purple), $0.7$  (blue), $0.3$ (light-blue) and $0.1$ (cyan), respectively.
}
\label{Fig:bl3thermal}
\end{figure}

We show, in Fig.~\ref{Fig:bl3thermal}, conditions
 to reproduce the observed DM relic abundance $\Omega h^2\simeq 0.1$ in a $(m_S - m_{Z^\prime})$ plane.
The purple, blue, light-blue and cyan correspond to $g_{(B-L)_3} = 1, 0.7, 0.3$ and $0.1$, respectively. 
One can easily find the case that $SS \rightarrow Z'Z'$ annihilation is available for $m_S > m_{Z'}$ as well as
 $Z'$ resonant enhanced regions along the line $m_S \simeq m_{Z'}/2$.

%

\subsection{Prospect for direct and indirect dark matter detection experiments}

\begin{figure}[htbp] 
 \centering
    \begin{tabular}{c}
 \begin{minipage}{0.48\hsize}
\centering
\includegraphics[width=8.0cm]{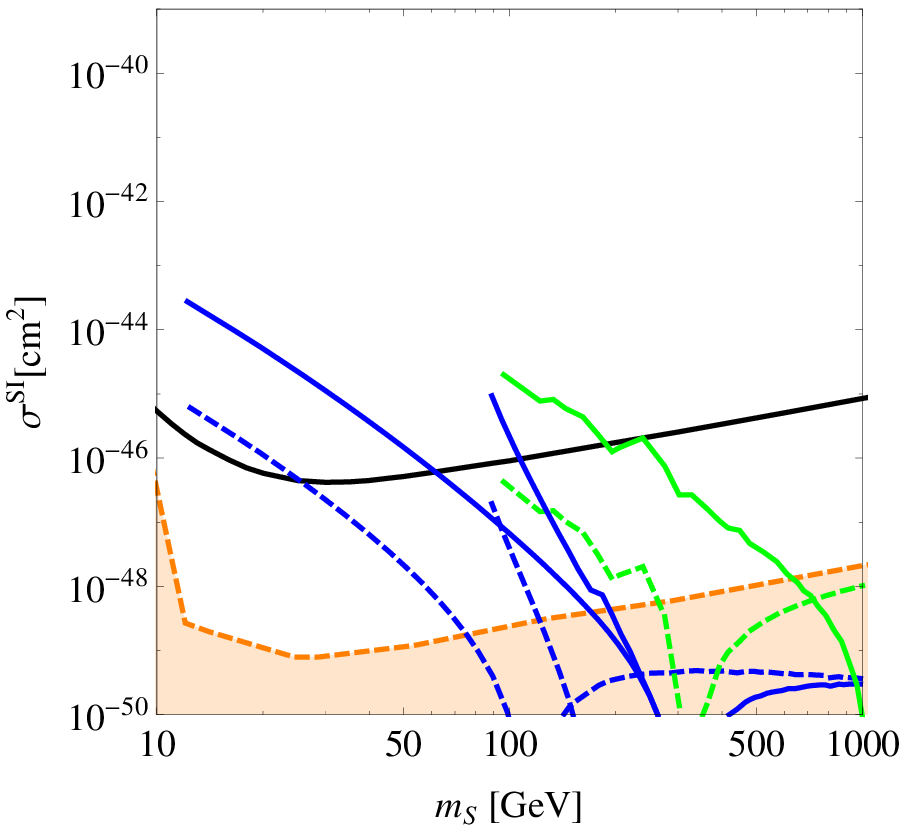}
 \end{minipage}
      \begin{minipage}{0.04\hsize}
        \hspace{2mm}
      \end{minipage}
 \begin{minipage}{0.48\hsize}
\centering
\includegraphics[width=8.0cm]{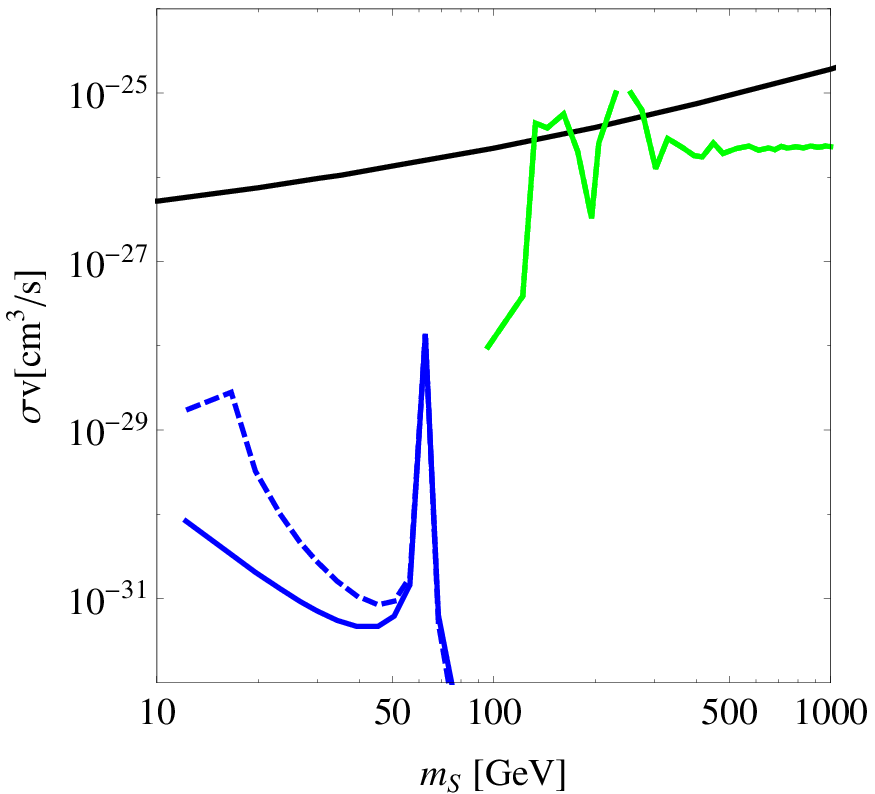}
 \end{minipage}   
 \end{tabular}
\caption{
The prediction of the WIMP-nucleon scattering cross section (Left) and
 the prediction of the present DM pair annihilation cross section (Right) for $g_{(B-L)_3}=0.3$. 
The shadings and black solid curves are same as in Fig.~\ref{Fig:bldetection}.
The blue and green curves correspond to the $Z^\prime$ funnel annihilation
 and annihilation through $SS \rightarrow Z'Z'$ channel, respectively, as in Fig.~\ref{Fig:bl3thermal}.
The solid (dashed) curve corresponds to $m_H = v_2/4 \,\, (v_2/10)$.
}
\label{Fig:bl3detection}
\end{figure}

\begin{figure}[htbp] 
 \centering
    \begin{tabular}{c}
 \begin{minipage}{0.48\hsize}
\centering
\includegraphics[width=8.0cm]{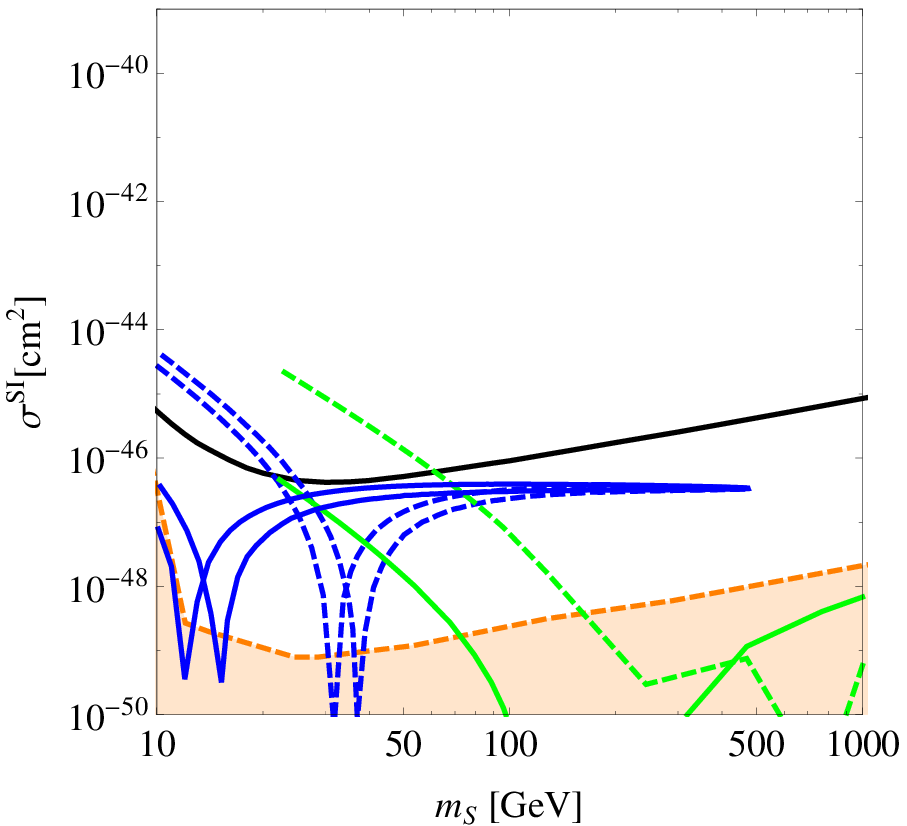}
 \end{minipage}
      \begin{minipage}{0.04\hsize}
        \hspace{2mm}
      \end{minipage}
 \begin{minipage}{0.48\hsize}
\centering
\includegraphics[width=8.0cm]{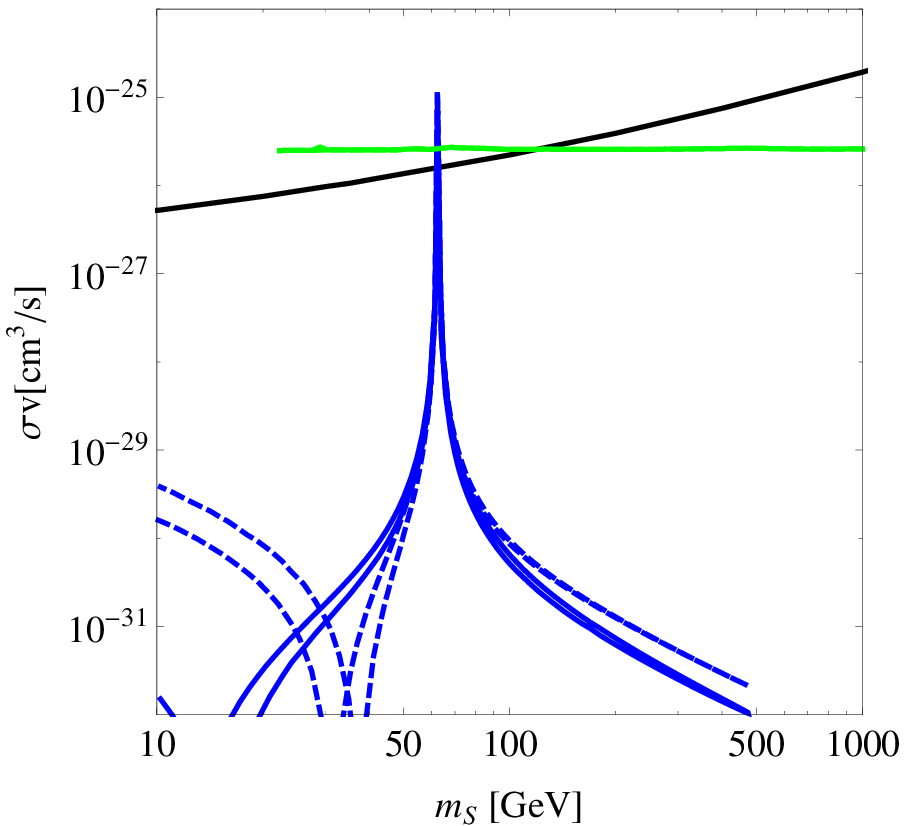}
 \end{minipage}   
 \end{tabular}
\caption{
Same as Fig.~\ref{Fig:bl3detection} for $g_{(B-L)_3}=0.03$.
}
\label{Fig:bl3(003)detection}
\end{figure}

The expected spin-independent cross section and the expected present DM pair annihilation cross section
 as a function of the DM mass satisfying the DM abundance are shown in Fig.~\ref{Fig:bl3detection} for $g_{(B-L)_3}=0.3$
 and in Fig.~\ref{Fig:bl3(003)detection} for $g_{(B-L)_3}=0.03$.
In both figures, blue curves are for $m_S < m_{Z'}$ while green curves are for $m_S > m_{Z'}$.
We can find that an expected present annihilation cross section is $\lesssim \mathcal{O}(10^{-30})$ cm$^3$/s
 for its wide mass range.

We may apply our scenario to explain an excess in $\gamma$-ray emission from the globular cluster 47 Tucanae,
 which could be interpreted as DM annihilation with the mass about $34$ GeV and
 the annihilation cross section of $6 \times 10^{-30}$ cm$^3$/s
 mainly into $b\bar{b}$ quarks~\cite{Brown:2018pwq,Lloyd:2018urr}.\footnote{
 Millisecond pulsars interpretation has also been pointed out~\cite{Bartels:2018qgr}.}
%

%

\section{$L_{\mu}-L_{\tau}$ Model}
\label{Sec:MuTau}

\subsection{$L_{\mu}-L_{\tau}$ seesaw model}

\begin{table}[t]
\begin{center}
\begin{tabular}{|c|ccc|c|}
\hline
      &  SU(3)$_c$  & SU(2)$_L$ & U(1)$_Y$ & U(1)$_{L_{\mu}-L_{\tau}}$  \\ 
\hline
$Q^{i}$ & {\bf 3 }        &  {\bf 2}         & $ 1/6$     &  $0 $   \\
$u^{i}_{R}$ & {\bf 3 }    &  {\bf 1}         & $ 2/3$     &  $0 $   \\
$d^{i}_{R}$ & {\bf 3 }    &  {\bf 1}         & $-1/3$     &  $0 $   \\
\hline
$L^{i}$        & {\bf 1 }    &  {\bf 2}       & $-1/2$     &  $0 $ for $i=1$,  \quad $1 $ for $i=2$,  \quad $-1 $ for $i=3$ \,   \\
$e^{i}_{R}$    & {\bf 1 }    &  {\bf 1}       & $-1$       &  $0 $ for $i=1$,  \quad $1 $ for $i=2$,  \quad $-1 $ for $i=3$ \,   \\
\hline
$\Phi$            & {\bf 1 }    &  {\bf 2}       & $ 1/2$    & $0 $   \\  
\hline
$N^{i}_{R}$    & {\bf 1 }    &  {\bf 1}       &$0$         &  $0 $ for $i=1$,  \quad $1 $ for $i=2$,  \quad $-1 $ for $i=3$ \,     \\
$\phi_1$       & {\bf 1 }    &  {\bf 1}       &$ 0$        & $ + 1 $  \\
$\phi_2$       & {\bf 1 }    &  {\bf 1}       &$ 0$        & $ + 2 $  \\ 
\hline
\end{tabular}
\end{center}
\caption{
The particle content of our $U(1)_{L_{\mu}-L_{\tau}}$ model. 
In addition to the SM particle content ($i=1,2,3$), three RH neutrinos  
  ($N_R^i$ ($i=1, 2, 3$)) and two $U(1)_{L_{\mu}-L_{\tau}}$ Higgs fields ($\phi_1$ and $\phi_2$) are introduced.   
}
\label{tableMuTau}
\end{table}

Finally, we consider another interesting anomaly free flavored extra $U(1)$ gauge symmetry,
 $U(1)_{L_{\mu}-L_{\tau}}$~\cite{He:1990pn,Foot:1990mn} .
This is interesting because the extra gauge boson may solve~\cite{Baek:2001kca,Ma:2001md}
 discrepancy between the experimental result and the SM prediction on
 the muon anomalous magnetic moment~\cite{Bennett:2006fi,Jegerlehner:2009ry,Davier:2010nc,Hagiwara:2011af}.
Two RH neutrinos ($N_R^2$ and $N_R^3$) are also charged under the $U(1)_{L_{\mu}-L_{\tau}}$ and
 two $U(1)_{L_{\mu}-L_{\tau}}$ Higgs fields ($\phi_1$ and $\phi_2$) are introduced. 
The particles and those charges are listed on Table~\ref{tableMuTau}.

\subsection{Thermal relic abundance}

\begin{figure}[htbp]
\centering
\includegraphics[clip,width=12.0cm]{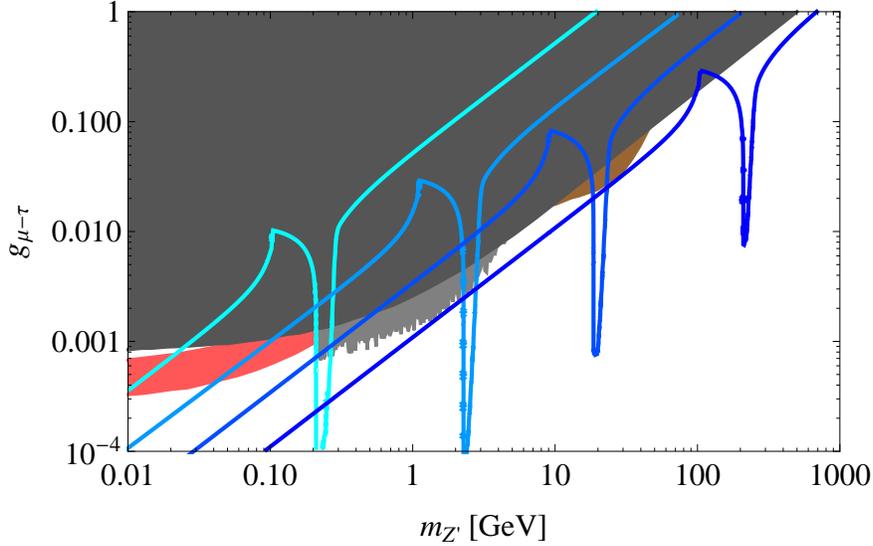}
\caption{
The required magnitude of the gauge coupling $g_{\mu-\tau}$
 to reproduce the observed DM relic abundance $\Omega h^2 \simeq 0.1$ for
$m_S=0.1$ GeV (turquoise), $1$ GeV (cyan), $10$ GeV (blue) and $100$ GeV (navy), respectively. 
}
\label{Fig:mt_thermal}
\end{figure}

We show, in Fig.~\ref{Fig:mt_thermal}, experimental constraints,
 a parameter region favored to solve the muon anomalous magnetic moment
 and the contours to reproduce the observed DM relic abundance $\Omega h^2\simeq 0.1$
 for various $m_S$ values with $m_P -m_S = 0.01 m_S$
 in ($m_{Z^\prime}, g_{\mu-\tau}$) plane, where $g_{\mu-\tau}$ is the $U(1)$ gauge coupling.
The light gray region is excluded by the \textit{BABAR} experiment~\cite{TheBABAR:2016rlg}.
The gray region is constrained by nonobservation of the neutrino trident
 processes~\cite{Altmannshofer:2014pba,Mishra:1991bv}.
The LHC bound has been studied in Refs.~\cite{Chun:2018ibr,Sirunyan:2018nnz,Drees:2018hhs} and
 will be exhibited as a brown-shaded region in the plot.
The vermilion region is favored to account for the discrepancy in the anomalous magnetic moment of
 muon~\cite{Baek:2001kca,Ma:2001md,Altmannshofer:2014pba}.
The big bang nucleosynthesis bound has been obtained
 as $m_{Z'} \gtrsim 5$ MeV by demanding $\Delta N_\mathrm{eff} < 0.7$~\cite{Kamada:2015era,Kamada:2018zxi}.
On the other hand, it is also recently pointed out that the ``Hubble tension'',
 which is the discrepancy between the values of the Hubble constant as determined from
 local measurements~\cite{Riess:2016jrr,Riess:2018byc} 
 and estimated from the temperature anisotropies of the cosmic microwave background~\cite{Aghanim:2018eyx},
 can be relaxed in the $U(1)_{L_{\mu}-L_{\tau}}$ model 
 by increasing the number of relativistic degrees of freedom as $\Delta N_\mathrm{eff}\simeq 0.2$~\cite{Escudero:2019gzq}.
The turquoise, cyan, blue and navy contours correspond to $m_S = 0.1, 1, 10$ and $100$ GeV, respectively. 
A sharp drop is due to the rapid annihilation by $s$-channel $Z'$ resonance pole. 
For a lighter mass region of $Z'$ as $m_{Z'} < m_S$,
 the annihilation mode $S S \rightarrow Z^\prime Z'$ is dominant.
As we will see in Fig.~\ref{Fig:mt_detection}, for a wide mass range of $m_S \lesssim 100$ GeV,
 the $m_{Z'} < m_S$ case is constrained by the Fermi-LAT bound on the present DM annihilation
 cross section $SS \rightarrow Z'Z'$ followed by the decay
 of $Z' \rightarrow \tau^+\tau^-, \mu^+\mu^-$ and $\nu\bar{\nu}$.
One exception is that, for a very small mass case, e.g., $m_S \simeq 0.1$ GeV drawn by the turquoise curve, 
 it is free from the constraint because $Z'$ cannot decay into charged leptons but do into only neutrinos. 
Remarkably, such a DM mass range with $m_{Z'}\simeq \mathcal{O}(0.01)$ GeV can solve
 the muon $g-2$ anomaly, explain the feature of high energy neutrino spectrum~\cite{Araki:2014ona,Araki:2015mya}
 measured by the IceCube Collaboration~\cite{Aartsen:2014gkd} as well as relax the Hubble tension.
In addition, a heavier $Z^\prime$ mass $> 10$ GeV in the $L_\mu-L_\tau$ model is also favored to solve
 the so-called $b \rightarrow s \mu^+\mu^-$ anomaly~\cite{Altmannshofer:2014cfa,Altmannshofer:2016jzy} reported by
 the LHCb~\cite{Aaij:2013qta,Aaij:2014ora,Aaij:2015esa,Aaij:2015oid} and the Belle~\cite{Abdesselam:2016llu} experiments.
In comparison to our scalar DM, $L_\mu-L_\tau$ charged fermion DM~\cite{Cirelli:2008pk,Baek:2008nz}
 with such a $Z^\prime$ mass suffers from the compatibility of the thermal abundance and the LHC bound,
 unless the charge is larger than $3$~\cite{Foldenauer:2018zrz}.
%

\subsection{Prospect for direct and indirect dark matter detection experiments}

\begin{figure}[htbp] 
 \centering
    \begin{tabular}{c}
 \begin{minipage}{0.47\hsize}
\centering
\includegraphics[width=8.0cm]{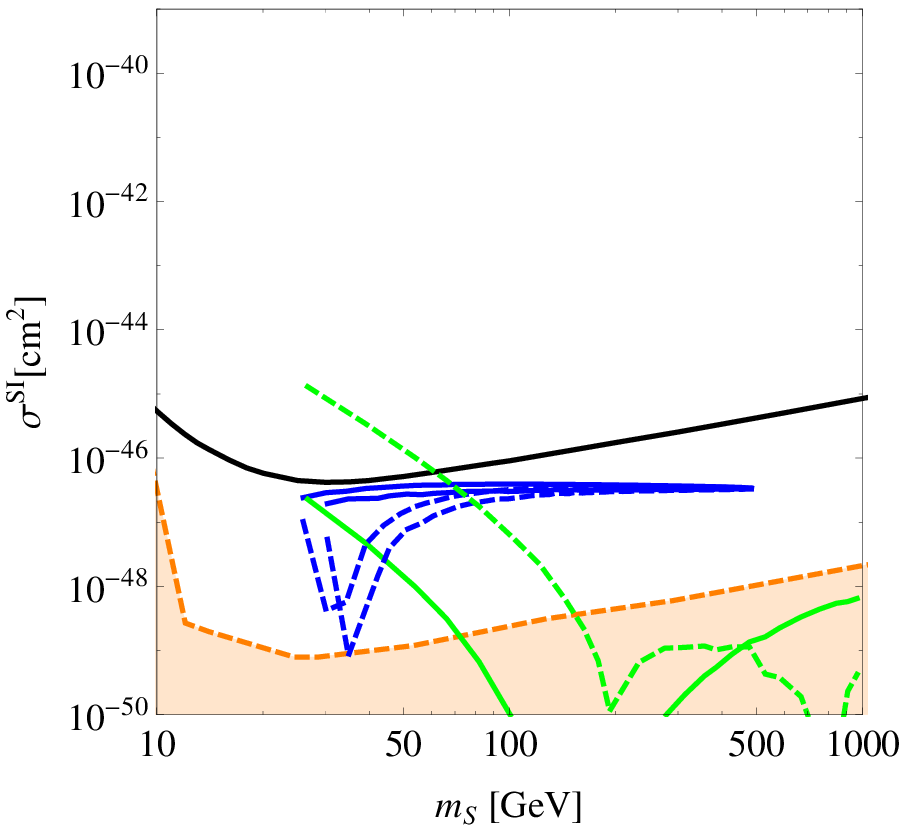}
 \end{minipage}
      \begin{minipage}{0.06\hsize}
        \hspace{2mm}
      \end{minipage}
 \begin{minipage}{0.47\hsize}
\centering
\includegraphics[width=8.0cm]{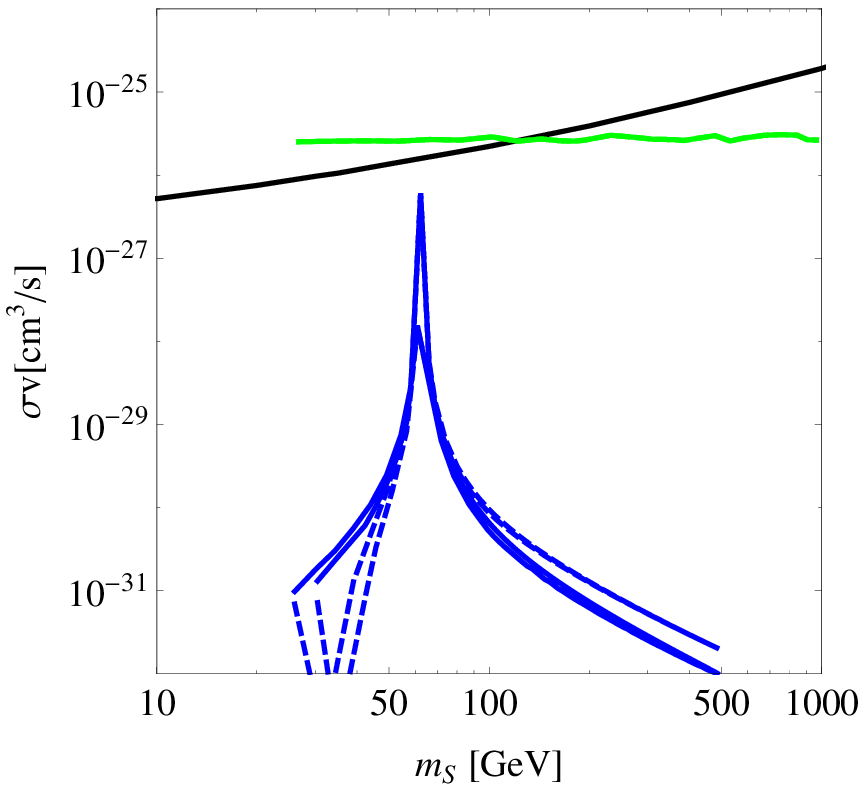}
 \end{minipage}   
 \end{tabular}
\caption{
Same as Fig.~\ref{Fig:bl3detection} for $g_{\mu-\tau}=0.03$.
The shadings and various curves are same as in Fig.~\ref{Fig:bl3detection}.
}
\label{Fig:mt_detection}
\end{figure}

We show in Fig.~\ref{Fig:mt_detection} the spin-independent cross section $\sigma^{\mathrm{SI}}$ (left) and 
 the prediction of the present DM pair annihilation cross section (right).
In the right panel, the green solid and dashed curves lie on each other.
In both figures, blue curves are for $m_S < m_{Z'}$ while green curves are for $m_S > m_{Z'}$.

The constraints~\cite{Boudaud:2016mos} and future prospects~\cite{Bartels:2017dpb,Dutra:2018gmv}
 on sub-GeV WIMP have been studied for annihilation into $e^- e^+$.
However, those cannot directly be applied to our case,
 since $S$ does not annihilate into $e^- e^+$ and 
 $Z'$ decays into only neutrinos as long as the mass of ${Z'}$ is lighter than
 the muon mass.\footnote{Through the kinetic mixing, $Z^\prime$ could decay into $e^- e^+$. 
Then, the constraint can be interpreted as just the upper bound on the mixing.
This constraint on the kinetic mixing will be studied elsewhere.}

%

\section{Summary}
\label{Sec:Summary} 

The current constrains by null results of DM direct detection experiments could
 impose stringent limits on the present DM annihilation cross section. 
From the null detection, scalar DM annihilating into $b\bar{b}$ through
 the Higgs boson exchange is naively expected to have the present annihilation cross section be
 smaller than $\mathcal{O}(10^{-31})$ cm$^3/$s for its wide mass range,
 which is much more stringent than the limit derived from Fermi-LAT experiment. 

We have investigated a possible realization of a thermal relic DM with the present annihilation
 cross section being very suppressed to be $\mathcal{O}(10^{-31})$ cm$^3$/s. 
One simple way is to introduce the $U(1)$ gauge interaction,
 where the thermal DM abundance is determined by coannihilation through the gauge interaction
 while the present annihilation is governed by Higgs bosons exchange processes. 
The VEV of the extra $U(1)$ breaking Higgs field generates a small mass splitting between DM
 and its coannihilating partner so that
 coannihilation becomes possible and the $Z'$-mediated scattering off with a nucleon in  direct DM search
 becomes irrelevant due to its inelastic nature.
We have examined three specific anomaly free models, $U(1)_{B-L}, U(1)_{(B-L)_3}$ and $U(1)_{L_\mu-L_\tau}$.
Only heavy WIMP is possible for the universal $U(1)_{B-L}$ model because the LHC bound on the $Z'$ boson is so stringent,
 while the weak scale mass or even lighter thermal WIMPs are possible for the other flavored $U(1)$ models. 
WIMPs on those models can be detected at future experiments and its hint might be already detected
 as a $\gamma$-ray excess in the globular cluster 47 Tucanae.
For $U(1)_{L_\mu-L_\tau}$ model, 
 a part of parameter region of $SS \rightarrow Z'Z'$ is going to be constrained.
A heavier mass region $m_{Z'} > m_S > 10$ GeV is interesting,
 because it can be compatible with solving the $b \rightarrow s \mu^+\mu^-$ anomaly.
In addition, a light mass region $m_S\simeq 0.1$ GeV with $m_{Z'}= 0.01-0.1$ GeV is particularly interesting,
 because the discrepancy of the muon anomalous magnetic moment as well as
 the Hubble tension can be relaxed.


\section*{Acknowledgments}
This work is supported in part by the U.S. DOE Grant No.~DE-SC0012447 (N.O.) and
KAKENHI Grants No.~19K03860, No.~19H05091 and No.~19K03865 (O.S.).

%

\appendix

\section{The decay width of $Z'$ boson}

We obtain the partial decay width of $Z'$ into $S$ and $P$ as
%
\begin{align}
\Gamma(Z'\rightarrow SP) = \frac{g'^2}{16 \pi} \frac{(m_{Z'}^2-(m_P-m_S)^2)^{3/2}(m_{Z'}^2-(m_P+m_S)^2)^{3/2} }{m_{Z'}^5}
\end{align}
 from the vertex (\ref{eq:Lag:gauge-DM-DM}).
Other partial decay widths for other channels can be found in Ref.~\cite{Carena:2004xs}.

\section{Annihilation cross section}
\label{App:amplitude}

\subsection{Annihilation into charged fermions}
%
%
\begin{align}
 & \overline{\left|{\cal M}(S P \rightarrow f\bar{f})\right|^2} \nonumber \\
 = & g'^4 \left( q^f\right)^2 N_c \frac{2}{m_{Z'}^4}\frac{1}{(s-m_{Z'}^2 )^2+(m_{Z'}\Gamma_{Z'} )^2} \nonumber \\
& \left( m_{Z'}^4 \left[(2 m_f^2 +m_S^2 +m_P^2)\left(s-2(m_P^2+m_S^2)\right) +(m_S^2-m_P^2)^2 \right]\right. \nonumber \\
 & \left. +(m_S^2-m_P^2)^2\left[ m_{Z'}^2 \left(2 m_f^2+m_P^2+m_S^2-s\right)+(m_S^2 +m_P^2 -2 m_f^2 )s-(m_S^2 -m_P^2)^2 \right] \right)
 \\
 \rightarrow & 
g'^4 \left( q^f \right)^2 N_c \frac{4(m_S^2+m_f^2)}{(s-m_{Z'}^2 )^2+(m_{Z'}\Gamma_{Z'} )^2}(s-4m_S^2) \qquad \mathrm{for} \qquad m_P\rightarrow m_S.
\end{align}

\subsection{Annihilation into neutrinos}
%
%
\begin{align}
 & \overline{\left|{\cal M}(S P \rightarrow \nu\bar{\nu})\right|^2} \nonumber \\
 = & g'^4 \left( q^f \right)^2 N_c \frac{2}{m_{Z'}^4}\frac{1}{(s-m_{Z'}^2 )^2+(m_{Z'}\Gamma_{Z'} )^2} \nonumber \\
& \left( m_{Z'}^4 \left[(-2 m_{\nu}^2 +m_S^2 +m_P^2)\left(s-2(m_P^2+m_S^2)\right) +(m_S^2-m_P^2)^2 \right]\right. \nonumber \\
 & \left. +(m_S^2-m_P^2)^2\left[ m_{Z'}^2 \left(- 2 m_{\nu}^2+m_P^2+m_S^2-s\right)+(m_S^2 +m_P^2 +2 m_{\nu}^2 )s-(m_S^2 -m_P^2)^2 \right] \right)
 \\
 \rightarrow & 
g'^4 \left( q^f \right)^2 N_c \frac{4(m_S^2-m_{\nu}^2)}{(s-m_{Z'}^2 )^2+(m_{Z'}\Gamma_{Z'} )^2}(s-4m_S^2) \qquad \mathrm{for} \qquad m_P\rightarrow m_S.
\end{align}

\subsection{Annihilation into $Z' Z'$}
%
%
\begin{align}
 & \int\frac{d\cos\theta}{2} \overline{\left|{\cal M}(S S \rightarrow Z' Z')\right|^2} \nonumber \\
 = & g'^4 \left( q^{\Phi} \right)^4 \left( \frac{4(16m_S^4+8m_S^2(m_{Z'}^2-2s)-3m_{Z'}^4+4 m_{Z'}^2s+s^2)}{(s-2m_{Z'}^2)\sqrt{(s-4m_S^2)(s-4m_{Z'}^2)}} \right. \nonumber \\
& \times\log\left( \frac{s-2m_{Z'}^2+\sqrt{(s-4m_S^2)(s-4m_{Z'}^2)}}{s-2m_{Z'}^2-\sqrt{(s-4m_S^2)(s-4m_{Z'}^2)}} \right)    \nonumber \\
& \left. + \frac{1}{m_{Z'}^4}\frac{32m_S^4m_{Z'}^4+m_S^2(-32m_{Z'}^6+20 m_{Z'}^4s-8m_{Z'}^4s^2+s^3)+m_{Z'}^4(6m_{Z'}^4-4m_{Z'}^2s+s^2)}{m_S^2(s-4m_{Z'}^2)+m_{Z'}^4} \right)
\end{align}
for the $m_P \rightarrow m_S$ limit.

%



\end{document}